# Using Crowdsourcing to Identify a Proxy of Socio-Economic status

Adil E. Rajput [1], Akila Sarirete [1], Tamer F. Desouky [2]

[1]College of Engineering, Effat University, Jeddah, KSA
[2] College of Humanities, Effat University, Jeddah, KSA
`aiilahi@effatuniversity.edu.sa`

**Abstract.** Social Media provides researchers with an unprecedented opportunity to gain insight into various facets of human life. Health practitioners put a great emphasis on pinpointing socioeconomic status (SES) of individuals as they can use to it to predict certain diseases. Crowdsourcing is a term coined that entails gathering intelligence from a user community online. In order to group the users online into communities, researchers have made use of hashtags that will cull the interest of a community of users. In this paper, we propose a mechanism to group a certain group of users based on their geographic background and build a corpus for such users. Specifically, we have looked at discussion forums for some vehicles where the site has established communities for different areas to air their grievances or sing the praises of the vehicle. From such a discussion, it was possible to glean the vocabulary that these group of users adheres to. We compared the corpus of different communities and noted the difference in the choice of language. This provided us with the groundwork for predicting the socio-economic status of such communities that can be particularly helpful to health practitioners and in turn used in smart cities to provide better services to the community members. More work is underway to take words and emojis out of vocablary(OOV) and assessing the average score as special cases.

## 1 Introduction

The social media platform presents an unprecedented opportunity for practitioners in the fields of medical and social sciences. It provides a forum for users to express their opinions in a disinhibited fashion as they need not disclose their real identity. While many social media platforms today do require the end users to confirm their real identity to a certain degree, the process is not guaranteed 100%. Furthermore, social media companies are bound by federal regulations to safeguard the real identity of the end user. The US presidential campaign of 2004 made the concept of online campaigning popular and propelled certain authors to study the effectiveness of such campaigns (Weinberg & William, 2006).

Researchers in social and medical sciences have started paying attention to looking at the plethora of data that is available to them. While the data available cannot be used to detect and diagnose the problems being faced by a certain individual per se, the data available can provide a basis for detecting various symptoms that can prove a harbinger for the onset of certain mental health issues (Rajput & Ahmed, 2018a). The techniques developed in the Natural Language Processing (NLP) domain can prove invaluable in processing, segmenting and clustering the available text data per the various segmentation techniques needed by the social and medical sciences practitioners. The choice of the corpus is one of the essential requirements to perform these series of steps. A corpus is defined as "A collection of naturally occurring text, chosen to characterize a state or variety of a language" (Schvaneveldt, Meyer, & Becker, 1976). Generally speaking, development of corpus entails looking at a text specific to the problem at hand and deriving keywords, bigrams and at times trigrams (two and three-word phrases) that are heavily used in a particular domain. As an example, authors in (Rajput & Ahmed, 2018b) argue for the need of establishing a corpus that would help mental health practitioners detect depression among users given a certain group. The authors looked at hashtag #depression on twitter and gleaned the keywords and established that such words were part of the vocabulary of depression patients (Rajput & Ahmed, 2018b). Once such a corpus is established, researchers would look at a random text and predict with a certain assurance whether the words used by a given person show the same frequency as those in the corpus.

One of the dimensions that mental health practitioners and sociologists look at is the socio-economic status (SES) of a given individual. The SES status is used to predict the potential issues the individual might face. As an example, the National Institute of Health (NIH) discusses the comorbidity of SES and alcoholism in Collins (2016). One of the key factors that determine the SES of an individual is the level of education and in turn, the writing style of an individual could be used as an indicator of their level of education. The Flesch-Kincaid test (Kincaid, Fishburne, Rogers & Chissom, 1975) developed a formula that combines the following ratios of a text to determine the grade level of the text at hand, as follows:

1. The ratio of total words to total sentences in a given text
2. The ratio of total syllables to total words in a given text

The grade level obtained from the above formula described above correlates directly to the school grade level of the text at hand. For example, a grade level of 12 indicates that a student in grade level 12 of school could comprehend the text at hand while a grade level of 14 would indicate that the text is written at the level of a student in the second year of University.

We looked at a public online discussion forum where users who bought a particular car posted their thoughts and impressions. The discussion forum was segmented already by various regions in the USA and Canada (e.g., Northeast, South, Western Ontario etc.). We scavenged the forum and ran various experiments to see whether individuals from different regions differ in Flesh-Kincaid grade level.

The paper will present the following:

1. Establish a corpus that represents the thoughts and expressions by users from different regions of the USA and Canada,
2. Compute the Flesh-Kincaid grade level score for the users and compute the average,
3. Compare the scores across the regions to see whether there is any difference.

## 2      Literature Review

Socioeconomic status (SES) is an important composite index used to measure the social standing of an individual or groups of individuals along socially constructed group identities or along geographic locations (Baker, 2014). It is typically composed of an individual's education, income, and occupation Education can be measured by a number of formal schooling years completed or along a Likert scale. Income, on the other hand, can be measured using objective data such as annual income or more subjective measures such as economic stress and the ability to pay immediate bills (Chen & Paterson, 2006). Finally, the occupation can be measured by job titles as well as standardized lists. Rank order of prestige is derived from opinion polls and other organizations. However, some researchers suggest alternative ways of measuring occupation as subjective in nature by allowing respondents to self-compare to others in a social hierarchy (Diemer et al., 2013).

Numerous fields in the social sciences use SES as a significant predictor of important variables of interest. For example, SES has been shown to have a medium to a strong relationship with academic achievement (White, 1982; Sirin, 2005) as well as causal effects on health (Adler & Ostrove, 1999), for instance, the association between SES and obesity is well-established. An update (McLaren, 2007) to the original review by Sobal & Stunkard (1989) has shown consistent results. Interestingly, SES has even been shown to be a predictor of brain function (specifically language and executive functions) (Hackman & Farah, 2009). From studies looking at child development (Bradley & Corwyn, 2002) to how minority groups experience and cope with stress (Gayman, Cislo, Goidel, & Ueno, 2014), SES is an important variable to consider.

Some studies use measures of SES as a proxy for individual characteristics when data is incomplete, and the validity of doing so has been questioned (Geronimus, Bound, & Neidert, 1996). An aggregate proxy (e.g., median household income) may inflate the effects shown in statistical analysis (Geronimus & Bound, 1998; Soobader, LeClere, Hadden, & Maury, 2001). However, our study is unique in that we do not use

SES as a proxy, but rather individual scores on a readability test as a reflection of educational attainment, which is a core factor in SES. By looking at a proxy measure of education via online posts, information can be collected to identify or even target these individuals in a more appropriate and customized manner.

One way to look at education is by measuring indirect constructs associated with it. As previously stated, SES has an effect on the human brain by affecting language attainment and expression as well as core executive functions (Hackman & Farah, 2009). Therefore, a measure of language expression, via writing, can potentially reflect an individual's SES. One common method used by the United States Military (Kincaid, Fishburne, Rogers, & Chissom, 1975) as well in word processing software is the Flesch–Kincaid Readability Tests (Stockmeyer, 2009). The Flesch Reading Ease Test rates text on a 100 point scale with higher scores indicating easier readability (Flesch, 1948). The Flesch-Kincaid Grade Level Test standardizes the score to the U.S. grade school levels (Kincaid, Fishburne, Rogers, & Chissom, 1975). For example, a score of 12.0 indicates that twelve graders can comprehend a given text. Some researchers have begun to use such measures in order to start adapting to the realm of big data (Flaounas et al., 2013). Much research is being conducted on readability with some measures superior to the Flesch-Kincaid test (Si, & Callan, 2001). However, even though progress has been made, some core problems still persist in terms of consistency (Mailloux, Johnson, Fisher, & Pettibone, 1995; Wang, Miller, Schmitt, & Wen, 2013).

Crowdsourcing techniques, on the other hand, are used in various scenarios to help quickly collect information about large groups of people in social networks. Wazn (2017) reviewed the definition of crowdsourcing, crowdsourcing taxonomies, crowdsourcing research, regulatory and ethical aspects, including some prominent examples of crowdsourcing. The author concluded that crowdsourcing has the potential to be extremely promising, in particular in healthcare as it has the ability to collect quickly and the information in a cost-effectively way. In order to group the users online into communities, researchers have made use of hashtags that will collect the interest of a community of users.

In this paper, we propose a mechanism to group a certain group of users based on their geographic background and build a corpus for such users. (Rajput &Ahmed 2018a) have presented a survey of various work that is done in clustering the mental health patients. In this study, , we will look at discussion forums for some vehicles where the site has established communities for different areas to air their grievances or sing the praises of the vehicle. From such a discussion, we can collect the vocabulary that these group of users adheres to. We compare the corpus of different communities and note the difference in the choice of language, therefore can draw some conclusion on the SES and some relationship with the academic achievement of individuals.

# 3 Experimental Setup

## 3.1 Assumptions

To evaluate the effectiveness of our corpus for social media, we focus on a public discussion forum of a given vehicle. Specifically, we capitalize on the fact that the discussion forum has been set up with various regions of the USA and Canada. The following assumptions hold true:

1. There is no way to confirm whether the users are posting in the right region. For example, a user from Northeast USA can post to the one in another region. While the users to report their respective city, there is no way for the forum to confirm whether the data is accurate.
2. We focus on corpora in the English language only and hence our results will only apply to the English speaking community.
3. We make sure that we do not count a duplicate post.
4. The posts by various users differ in size and we do not normalize the size. Rather, we compute the Flesch-Kincaid measure.

## 3.2 Data Sources and Data Gathering

One of the biggest challenges when gathering data is ensuring the legality of using the data (Youyou, Kosinski, & Stillwell, 2015). The data we gather comes from a public source. Various vehicle manufacturers have forums on the net for their customer to discuss their issues and help them monitor finds out potential problems so they can be addressed effectively. We do not store any user credentials and anonymize them by giving fictitious identifiers. Moreover, we use the following open source packages:

1. textstat package
2. beautifulsoup package

## 3.3 Pre-processing and Processing Data

We followed the following steps when implementing this project:

1. We use the built-in 'urllib' functionality of python to handle all url functionality;
2. We use the beautifulsoup package to scrape the data that is present in the forum;
3. We anonymize the user data to ensure that the privacy of individuals is not violated (even though the data is public);
4. We compute the Flesch-Kincaid grade level for each text within a forum;
5. We compute the average of each region and store it in the database.

We implemented the above on a standard Dell running Ubuntu Linux and Python3 program with a 16G RAM. Given that we did not have any performance requirements, the program can be ported to any platform that supports Python3.

### 3.4 Evaluation

After setting up the experiment, we ran the following experiments:

1. We computed the individual score for each post;
2. We computed the Flesch-Kincaid grade test for each post;
3. We tabulated the results and computed the average for each region and the standard deviation for each region;
4. A difference of 1 on the results meant a whole school grade difference so it was deemed significant.

## 4 Results and Discussion

We present the tabulated results below.

Table 1. Flesch-Kincaid grade level for each of the North American regions.

| Region | Total Number of | Average Flesch_Kincaid Score | Standard Deviation |
|---|---|---|---|
| North east | 66 | 6.6621212 | 3.072284 |
| Mid Atlantic | 101 | 6.514851 | 3.598927 |
| South | 22 | 8.409091 | 3.459412 |
| Midwest | 92 | 6.695652 | 3.440991 |
| Southwest | 41 | 7.317073 | 4.468999 |
| Northwest | 12 | 7.5 | 2.713602 |
| West | 36 | 5.583333 | 2.465476 |
| Canada-East | 56 | 6.5 | 3.092513 |
| Canada-West | 5 | 12.4 | 2.880972 |

Note the following:

1. Given that the number of posts in various regions are different, we use both the average and standard deviation to compare the results. As we can see from table 1 the Northeast, Mid-Atlantic, South, Midwest, and Canada-East regions have similar standard deviations. The total number of posts range from 22-101.

2. The actual data for each of the regions mentioned in 1 above along with Southwest regions contain 10-15% posts with a score of zero. Upon manual inspection, the authors noted that such posts either contained emojis or acronyms used on internet blogs such as "lol" – "laugh out loud".

3. While the Canada-West region shows the highest Flesch-Kincaid grade level, we ignore the results as the number of posts in that region are only 5.

4. The South and Northwest regions showed the highest average Flesch-Kincaid grade levels of 8.5 and 7.5 respectively. The Southwest region was not that behind with an average score of 7.3.

5. The Northeast, MidAtlantic. Midwest and Canada-East regions have similar average Flesch-Kincaid grade levels around 6.5.

Looking at point 5, we note that these regions usually contained more than one metropolitan area and the vehicle being discussed at hand could be owned by people with less education level as these regions are characterized by higher wages. The South and Southwest regions discussed in point 4 are characterized by wages that are less compared to the regions discussed in 5. Thus, people posting in regions discussed in points 4 show a writing level much less than this discussed in 4 and combining it with the average income could give us valuable insights into the SES of the people buying this vehicle.

## 5 Conclusion and Future Work

In this paper, our goal was to establish a framework through which we can start gathering information that can give researchers clues about the SES status of the people posting on various sites including social media. Such status can be helpful to medical and social scientists, practitioners and researchers. We established a corpus using the posts on a forum that was designed for a particular vehicle. The vehicle manufacturer divided the forum into six regions into the US and two regions in Canada. After gathering the posts, we computed the average Flesch-Kincaid grade level for each of the regions and compared the regions. We found that the South and Southwest regions showed the highest average grade level (we ignored the Canada-West region as it only had 5 posts).

We intend to do the following in the future:

1. Refine the work by considering Out of Vocabulary (OOV) words and emojis and evaluating the average score by handling them as special cases;

2. Looking at other forums that have a huge number of posts;

3. Computing a similar score for Twitter posts.